\documentclass[pra,twocolumn,showpacs,amsmath,amssymb,superscriptaddress,notitlepage]{revtex4-1}
\usepackage{psfrag,graphicx}
\usepackage{amsfonts,amssymb,amsmath}      
\usepackage{graphicx}
\usepackage{dcolumn}
\usepackage{bm}
\usepackage{float}
\usepackage{hyperref}
\usepackage{accents}
\usepackage{bbding}
\usepackage{color,soul}
\usepackage{epstopdf}
\usepackage{footnotehyper}

\newcommand{\erf}[1]{Eq.~(\ref{#1})}
\newcommand{\beq}{\begin{equation}}
\newcommand{\eeq}{\end{equation}}

\newcommand{\erfs}[2]{Eqs.~(\ref{#1})--(\ref{#2})}

\newcommand{\dg}{^\dagger}

\newcommand{\smallfrac}[2]{\mbox{$\frac{#1}{#2}$}}

\newcommand{\half}{\smallfrac{1}{2}}
\newcommand{\bra}[1]{\langle{#1}|}
\newcommand{\ket}[1]{|{#1}\rangle}

\newcommand{\Tr}{\text{Tr}}

\newcommand{\s}[1]{\hat{\sigma}_{#1}}
\renewcommand{\c}{_{{\rm O}_\Xi}}

\newcommand{\dd}{{\rm d}}

\newcommand{\past}[1]{\overleftarrow{#1}}%\overleftarrow{#1}}
\newcommand{\fut}[1]{\overrightarrow{#1}}
\newcommand{\both}[1]{\overleftrightarrow{#1}}
\newcommand{\fil}{_{\text F}}
\newcommand{\rfil}{_{\text R}}
\newcommand{\sm}{_{\text S}}

\newcommand{\god}{_{\text T}}%{^{\rm true}}%{_{\text G}}
\newcommand{\lust}{^{\rm L}}

\newcommand{\estrho}{\check{\rho}}

\definecolor{nblue}{rgb}{0.06,0.3,0.73}%229 11R, 61G, 145B
\definecolor{nblack}{rgb}{0,0,0}
\definecolor{nred}{rgb}{0.9,0.1,0.1}
\definecolor{nmagenta}{rgb}{0.7,0.0,0.3}

\newcommand{\blk}{\color{nblack}}

\begin{document}

\title{Quantum state smoothing as an optimal estimation problem \\ with three different cost functions}
\author{Kiarn T. Laverick}
\affiliation{Centre for Quantum Computation and Communication Technology 
(Australian Research Council), \\ Centre for Quantum Dynamics, Griffith University, Nathan, Queensland 
4111, Australia\\}
\author{Ivonne Guevara}
\affiliation{Centre for Quantum Computation and Communication Technology 
(Australian Research Council), \\ Centre for Quantum Dynamics, Griffith University, Nathan, Queensland 
4111, Australia\\}
\author{Howard M. Wiseman}
\affiliation{Centre for Quantum Computation and Communication Technology 
(Australian Research Council), \\ Centre for Quantum Dynamics, Griffith University, Nathan, Queensland 
4111, Australia\\}

\date{\today}
\begin{abstract}
Quantum state smoothing is a technique to estimate an unknown true state of an open quantum 
system based on partial measurement information both prior and posterior to the time of interest. In this paper, we 
show that the smoothed quantum state is an optimal state estimator; that is, it minimizes a 
risk (expected cost) function. Specifically, we show that the smoothed quantum state is optimal with respect to two cost functions: the trace-square 
deviation from and the relative entropy to the unknown true state. However, when we consider a related risk 
function, the linear infidelity, we find, contrary to what one might expect, that the smoothed state is not optimal. For 
this case, we derive the optimal state estimator, which we call the lustrated smoothed state. It is a pure state, the 
eigenstate of the smoothed quantum state with the largest eigenvalue. 
\end{abstract}
\maketitle

\section{Introduction}
Estimating the state of an open quantum system based on measurement information is an important task in 
quantum information science. Quantum trajectory theory, also referred to as quantum state 
filtering \cite{Bel87,Bel92,WisMil10}, utilises a continuous-in-time past measurement record $\past{\rm O}$ to 
estimate the quantum state of a single system at time $\tau$. One might think that one could obtain a 
more accurate estimate of the quantum state by conditioning on the {\em past-future} measurement record 
$\both{\rm O}$, that is, the measurement record both prior and posterior to the time of interest, $\tau$. However, 
due to the non-commutative nature of quantum states and operators, conditioning the estimate on future 
information is not straightforward in quantum mechanics, unlike in classical mechanics where the technique of 
smoothing is standard \cite{Weinert01, Hay01, vanTrees13, BroHwa12, Einicke12, Fri12, Sarkka13}. 
Consequently, utilising past-future measurement information in quantum systems has attracted great 
interest and many smoothing-like formalisms have been proposed
\cite{ABL64,AAV88,Tsa09a,Tsa09b,Cha13,GJM13,GueWis15,Ohki15,Ohki19,Tsa19}.
Here we are concerned with approach that is guaranteed to yield a valid quantum state 
as its estimate, the quantum state smoothing theory \cite{GueWis15,LCW19,CGW19,LCW-PRA21}. 

The quantum state smoothing formalism is as follows. Consider an open quantum system coupled to 
two baths (which could represent sets of baths). An observer, say Alice, monitors one bath and 
constructs a measurement record ${\rm O}$, called the `observed' record. The other bath, that is unobserved 
by Alice, is monitored by a second (perhaps hypothetical) observer, Bob, who also constructs a measurement 
record ${\rm U}$, called the `unobserved' record. Now, if Alice were able to condition her estimate of the 
quantum state on both the past observed and unobserved measurement she would have effectively 
performed a perfect measurement on the system and Alice's estimated state would be the true state of the system 
$\rho\god:=\rho_{\past{\rm O}\past{\rm U}}$ which can be assumed to be pure (it will be as long as the initial state 
is pure). Note, it is not a necessary 
requirement for Alice's and Bob's records together to constitute a perfect monitoring of the quantum system or for 
the true state to be pure, but merely a convenient assumption to make when introducing the idea of quantum state 
smoothing. But since Alice does not have access to the unobserved measurement record and she cannot know the 
true quantum state. However, Alice can calculate a Bayesian estimate of the true state, $\rho\c$, 
conditioned on her observed measurement record ${\rm O}_\Xi$ by averaging over all possible unobserved 
measurement records with the appropriate probabilities, i.e.,
\beq\label{Gen-sm}
\rho\c = \mathbb{E}_{\past{\rm U}|{\rm O}_\Xi}\{\rho\god\} 
\equiv \sum_{\past{\rm U}} \wp(\past{\rm U}|{\rm O}_\Xi) \rho_{\past{\rm O}\past{\rm U}}\,,
\eeq
where $\mathbb{E}\{\bullet\}$ denotes an ensemble average. For a filtered estimate of the quantum state, 
the past observed record is used (${\rm O}_\Xi = \past{\rm O}$) and $\rho_{\past{\rm O}} = \rho\fil$. For a 
smoothed estimate of the quantum state, the past-future observed record is used 
(${\rm O}_\Xi = \both{\rm O}$) and we define $\rho_{\both{\rm O}} = \rho\sm$.

Since its conception \cite{GueWis15}, various properties of and scenarios for the smoothed quantum state have 
been studied, providing considerable insight into the theory 
\cite{LCW19,CGW19,LCW-QS20,LCW-PRA21,Laverick21}. 
However, these works did not address one question: is the smoothed quantum state an 
optimal estimator of the true state and if so, in what sense? Here, `optimal' means minimizing a risk 
function. The risk function is the appropriately conditioned expectation for a cost function 
${\cal C}(\estrho,\rho\god)$, a measure comparing the estimate state $\estrho$ to the true state. That is, 
${\cal R}\c(\estrho) = \mathbb{E}_{\past{\rm U}|{\rm O}_\Xi}\{{\cal C}(\estrho,\rho\god)\}$. In this paper, 
we show for the trace-square deviation from the true state (Sec.~\ref{Sec-TrSD}) and the relative entropy 
with the true state (Sec.~\ref{Sec-RE}) as cost functions, the conditioned state (\ref{Gen-sm}) is the optimal 
estimator. Furthermore, for each cost function, we show that the risk function of the conditioned state 
$\rho_{{\rm O}_\Xi}$
reduces to simple measures involving only $\rho_{{\rm O}_\Xi}$. When we consider a cost function closely 
related to the trace-square deviation, the linear infidelity with the true state (Sec.~\ref{Sec-LI}), we find, 
somewhat counter-intuitively, that $\rho_{{\rm O}_\Xi}$ is not the optimal estimator. Rather, the {\em lustrated} 
conditioned state, a pure state corresponding to the largest-eigenvalue of $\rho_{{\rm O}_\Xi}$, is the optimal 
estimator. Finally, we derive upper and lower bounds on the risk function of the lustrated state for both the trace-
square deviation and linear infidelity risk function.

\section{Cost Function: Trace-Square Deviation}
\label{Sec-TrSD}
In this section, we will focus on the trace-square deviation (TrSD) from the true state, a distance measure 
between two quantum states, as the cost function of interest, i.e., 
${\cal C}(\estrho,\rho\god) = \Tr[(\estrho - \rho\god)^2]$. 
This means that the risk function for a given estimate $\estrho$ of the true quantum state is 
\beq
\begin{split}
\label{TrSD-Risk}
{\cal R}^{\rm TrSD}\c(\estrho) &= \mathbb{E}_{\past{\rm U}|{\rm O}_\Xi}\{\Tr[(\estrho - \rho\god)^2]\} \\ 
&= \mathbb{E}_{\past{\rm U}|{\rm O}_\Xi}\{P(\rho\god) - 2L(\estrho,\rho\god) + P(\estrho)\},
\end{split}
\eeq
where the purity is defined as 
\beq
P(\rho) = \Tr[\rho^2]\,,
\eeq 
and the linear fidelity \cite{Audenaert13} is 
\beq\label{LF}
L(\rho,\sigma) = \Tr[\rho\sigma]\,.
\eeq
Notice that here we have not restricted our discussion to smoothed estimates 
(${\rm O}_\Xi = \both{\rm O}$), but are also allowing for a filtered estimate 
(${\rm O}_\Xi = \past{\rm O}$). 

To see that the conditioned state, \erf{Gen-sm}, is the estimator that minimizes \erf{TrSD-Risk}, we will 
show that any other estimator $\rho' = \rho\c +\hat O$ for any traceless Hermitian operator $\hat O \ne 0$ 
is suboptimal. That is, the risk function for $\rho'$ is strictly greater than the risk function for 
$\rho\c$. By substituting $\rho'$ into the risk function we obtain 
\beq
\begin{split}
{\cal R}^{\rm TrSD}\c(\rho') &= \mathbb{E}_{\past{\rm U}|{\rm O}_\Xi}\{\Tr[(\rho' - \rho\god)^2]\}\\
&= {\cal R}^{\rm TrSD}\c(\rho\c) + 2\Tr[\hat{O}\rho\c] \\ &\qquad- 2\mathbb{E}_{\past{\rm U}|{\rm O}_\Xi}
\{\Tr[\hat{O} \rho\god]\} + \Tr[\hat{O}^2]\\
& = {\cal R}^{\rm TrSD}\c(\rho\c) + \Tr[\hat{O}^2]\\
&> {\cal R}^{\rm TrSD}\c(\rho\c)\,, 
\end{split}
\eeq
where we have used the cyclic and linear nature of the trace and the fact that $\hat O$ has real eigenvalues.

Interestingly, we can show that the expected linear 
fidelity between the conditioned state $\rho\c$ and the true state is equal to the purity of the conditioned 
state. That is, 
\beq\label{F=P}
\mathbb{E}_{\past{\rm U}|{\rm O}_\Xi}\{L(\rho\c,\rho\god)\} = 
\Tr[\rho\c\mathbb{E}_{\past{\rm U}|{\rm O}_\Xi}\{\rho\god\}] = P(\rho\c)\,.
\eeq
This equality also holds for the Jozsa fidelity \cite{Jozsa94}, 
$F(\rho\c,\rho\god) = \left(\Tr\left[\sqrt{\sqrt{\rho\god}\rho\c\sqrt{\rho\god}}\,\right]\right)^2$, provided that the true 
state is pure as, in this case, the Jozsa fidelity is equal to the linear fidelity.
Previously \cite{GueWis15,CGW19}, \erf{F=P} was proven only when also averaging both sides over the 
observed record ${\rm O}_\Xi$. (Note, in \cite{CGW19}, the equality between the Jozsa 
fidelity and the purity is typeset incorrectly; in Eq. (9), the left-hand side should also average over the past 
unobserved measurement record.) The risk function, \erf{TrSD-Risk}, for the optimal estimator 
$\estrho = \rho\c$, assuming a pure true state, thus reduces to the impurity of the conditioned state:
\beq\label{simp_risk}
{\cal R}^{\rm TrSD}\c(\rho\c) = 1 - P(\rho\c)\,.
\eeq

To verify \erf{F=P}, we will consider a physical example. The system is a two-level system, with a driving 
Hamiltonian $\hat{H} = (\Omega/2)\s{x}$ and radiative 
damping described by a Lindblad operator $\sqrt{\Gamma}\s{-}$, where $\Omega$ is the Rabi frequency 
and $\Gamma$ is the damping rate. Alice monitors a 
fraction $\eta$ of the output fluorescence using Y-homodyne detection so that her Lindblad operator can be written
$\hat{c}_\phi = \sqrt{\Gamma\eta}e^{-i\pi/2}\s{-}$, defined so that her photocurrent is 
$\dd J_\phi = \Tr[\hat{c}_\phi\rho + \rho\hat{c}_\phi\dg] + \dd W$, where 
$\dd W$ is a Wiener increment satisfying 
\beq\label{innovation}
\mathbb{E}\{\dd W\} = 0\qquad {\rm and} \qquad\mathbb{E}\{\dd W^2\} = \dd t. 
\eeq
The remaining $1-\eta$ fraction of the output is monitored by Bob using photodetection, with Lindblad operator 
$\hat{c}_N = \sqrt{\Gamma(1-\eta)}\s{-}$ defined so that the average rate of his jumps is 
${\mathbb E}\{\dd N\}/\dd t = \Tr[\hat{c}_N\dg\hat{c}_N \rho]$. Note that Alice and Bob's measurements collectively 
constitute a perfect measurement so that the true state is pure. We can describe the evolution of the true quantum 
state with the following stochastic master equation \cite{WisMil10}
\beq\label{SME-God}
\begin{split}
\dd \rho\god  = {\cal G}[\hat{c}_N]\rho&\dd N -i[\hat H,\rho\god]\dd t 
- {\cal H}[\half \hat{c}_N\dg \hat{c}_N]\rho\god\dd t 
\\&+ {\cal D}[\hat{c}_\phi]\rho\god\dd t + {\cal H}[\hat{c}_\phi]\rho\god \dd W\god\,,
\end{split}
\eeq
with initial condition chosen to be the ground state $\rho(t_0) = \ket{0}\bra{0}$.
The superoperators are defined as ${\cal G}[\hat{c}]\rho = \hat{c}\rho\hat{c}\dg/\Tr[\hat{c}\rho\hat{c}\dg] - \rho$, 
${\cal D}[\hat{c}]\rho = \hat{c}\rho\hat{c}\dg - \{\hat{c}\dg\hat{c},\rho\}/2$, and
${\cal H}[\hat{c}]\rho = \hat{c}\rho + \rho\hat{c}\dg - \Tr[\hat{c}\rho + \rho\hat{c}\dg]\rho$. Here, the vector of (true)
innovations $\dd W\god = \dd J_\phi - \Tr[\hat{c}_\phi \rho\god + \rho\god \hat{c}_\phi\dg]\dd t$ satisfies similar 
properties to those in \erf{innovation}. 

For the details of how the true state $\rho\god$, filtered state $\rho\fil$, the smoothed state $\rho\sm$ and 
the ensemble averages $\mathbb{E}_{\past{\rm U}|{\rm O}_\Xi}\{\bullet\}$ are computed, see Appendix 
\ref{App-code}. In Fig.~\ref{Fig1}, we can see the convergence of 
$\mathbb{E}_{\past{\rm U}|{\rm O}_\Xi}\{F(\rho\c,\rho\god)\}$ to the purity $P(\rho\c)$ as the 
number of unobserved trajectories averaged over increases, for both the filtered state and the smoothed state. 
When a large number ensemble is used, the two are almost indistinguishable. 

\begin{figure*}[t!]
\includegraphics[scale=0.51]{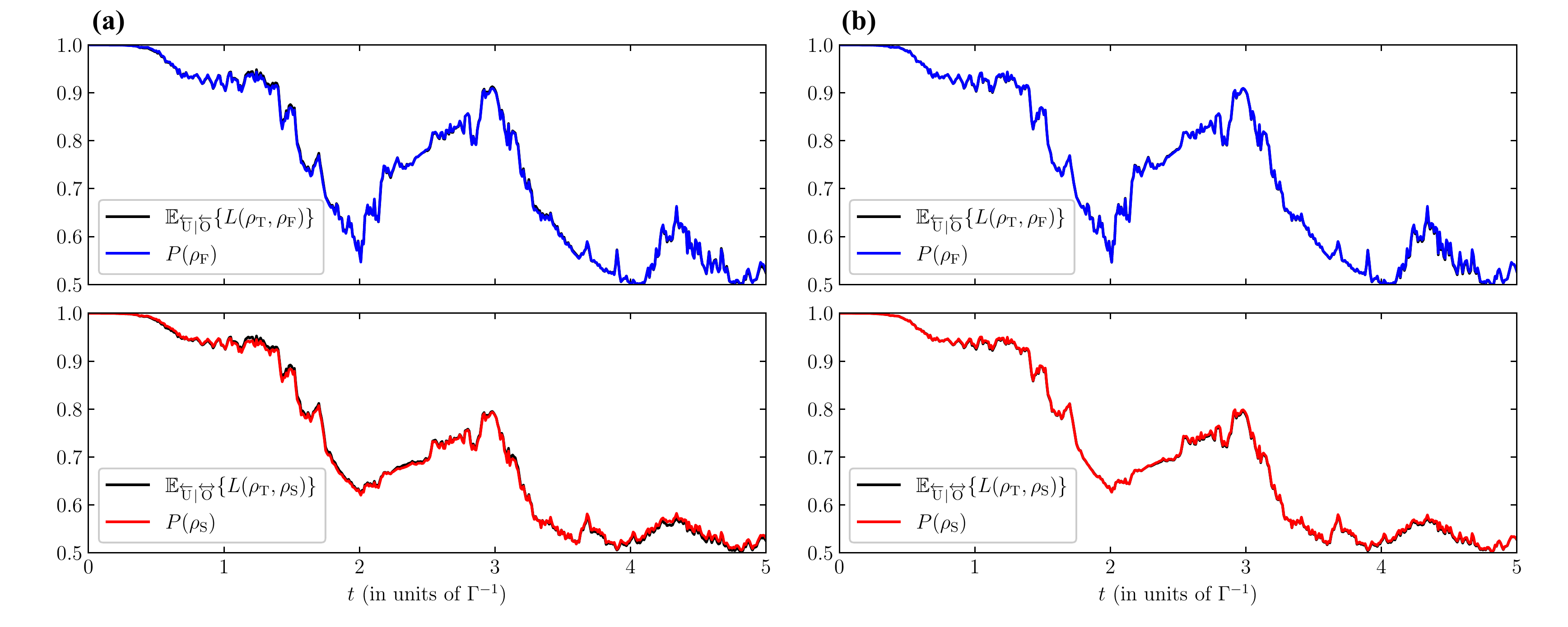}
\caption{The ensemble average of the linear fidelity between the conditioned state and the true state, and the purity 
of the conditioned state, for the quantum system of \erf{SME-God}, where Alice and Bob are monitoring the same 
channel with equal measurement efficiency ($\eta = 0.5$) using Y-homodyne detection and photodetection, 
respectively. To show the convergence of the average linear fidelity to the purity we have computed the averages 
using a small number of unobserved trajectories and also a larger number. Specifically, in the plots on the left-hand 
side, the ensemble averages for the smoothed quantum state and the average linear fidelity have been taken over 
$M = 2\times 10^3$ and $N = 3\times 10^3$ unobserved jump trajectories, respectively, while the corresponding 
figures for the right-hand side are $M = 2\times 10^4$ and $N = 3\times 10^4$. Note, 
the observed record is fixed in all cases and the ensemble of unobserved records used to generate the 
smoothed state and to average the linear fidelity are generated independently. As we can see, for both the filtered 
state (top graphs) and the smoothed state (bottom graphs), \erf{F=P} holds. Here, $\hbar = 1$, 
$\Omega = 3\Gamma$ and $\eta = 0.5$. }
\label{Fig1}
\end{figure*}

\section{Cost Function: Relative Entropy}
\label{Sec-RE}
We now move our attention to another cost function, the relative entropy with the true state. {\blk The consideration of this cost function in stems from the work in Ref.~\cite{GarCam19}, where the authors derived bounds on the average relative entropy between the state obtained by an omnicient observer (the true state) and either an ignorant (one who has no measurement information) or partially ignorant (one who has a fraction of the measurement output). The key difference here is that we are considering the optimal estimator that minimizes the (conditional) averarge of the relative entropy, as opposed to finding the upper and lower bounds. Furthermore, we consider a more general setting, allowing the omniscient observer to perform a different measurement (Bob's measurement in our formulation) on the remaining portion of the output to the partially ignorant observer (Alice in our formulation).}

The relative entropy is
defined as \cite{Ved02, Ruskai02, Audenaert13}
\beq\label{Rel-Ent}
S(\rho||\sigma) = \Tr[\rho\log\rho] - \Tr[\rho\log\sigma]\,.
\eeq 
$S(\rho||\sigma)$ is a measure of state distinguishability between state $\rho$ and $\sigma$, more specifically, it is 
akin to the likelihood that state $\sigma$ will not be confused with state $\rho$, where a relative entropy of zero 
corresponds to completely indistinguishable states. Note, unlike a lot of other cost functions, like the TrSD, 
the relative entropy is not symmetric in its arguments. 
%One can understand this asymmetry by first thinking of the von Neumann entropy defined as $S(\rho) = -\ex{\log\rho}$. The relative entropy is then simply the distance between the entropy of the actual state and the entropy of the state $\sigma$ that one has thinks the system is in, hence why the average is taken over the actual state of the system $\rho$ in the second term of \erf{Rel-Ent}. 
Consequently, the estimation task is to find the state that minimizes the risk function
\beq\label{RE=S}
{\cal R}^{\rm RE}\c(\estrho) = \mathbb{E}_{\past{\rm U}|{\rm O}_\Xi}\{S(\rho\god||\estrho)\}\,.
\eeq
  
Once again, the state that minimizes this risk function is the usual conditioned state. We 
can show the optimality of this state using the same method as before, that is, any state $\rho' \ne \rho\c$ 
will be suboptimal. Substituting $\rho'$ into the risk function we get
\beq
\begin{split}
{\cal R}^{\rm RE}\c(\rho') &= \mathbb{E}_{\past{\rm U}|{\rm O}_\Xi}\{\Tr[\rho\god\log\rho\god] - 
\Tr[\rho\god\log\rho']\}\\
& = \mathbb{E}_{\past{\rm U}|{\rm O}_\Xi}\{\Tr[\rho\god\log\rho\god] - \Tr[\rho\c\log\rho\c]\} \\&+ 
\mathbb{E}_{\past{\rm U}|{\rm O}_\Xi}\{\Tr[\rho\c\log\rho\c] - \Tr[\rho\god\log\rho'] \}\,.
\end{split}
\eeq
Remembering that $\rho\c = \mathbb{E}_{\past{\rm U}|{\rm O}_\Xi}\{\rho\god\}$, the risk function 
becomes
\beq
\begin{split}
{\cal R}\c^{\rm RE}(\rho') &= \mathbb{E}_{\past{\rm U}|{\rm O}_\Xi}\{S(\rho\god||\rho\c)\} + S(\rho\c||\rho') \\
&> \mathbb{E}_{\past{\rm U}|{\rm O}_\Xi}\{S(\rho\god||\rho\c)\}\,,
\end{split}
\eeq 
where the inequality results from the fact that the relative entropy is non-negative and 
is saturated only when $\rho' = \rho\c$ \cite{Ruskai02}. 

\begin{figure}[t]
\includegraphics[scale=0.53]{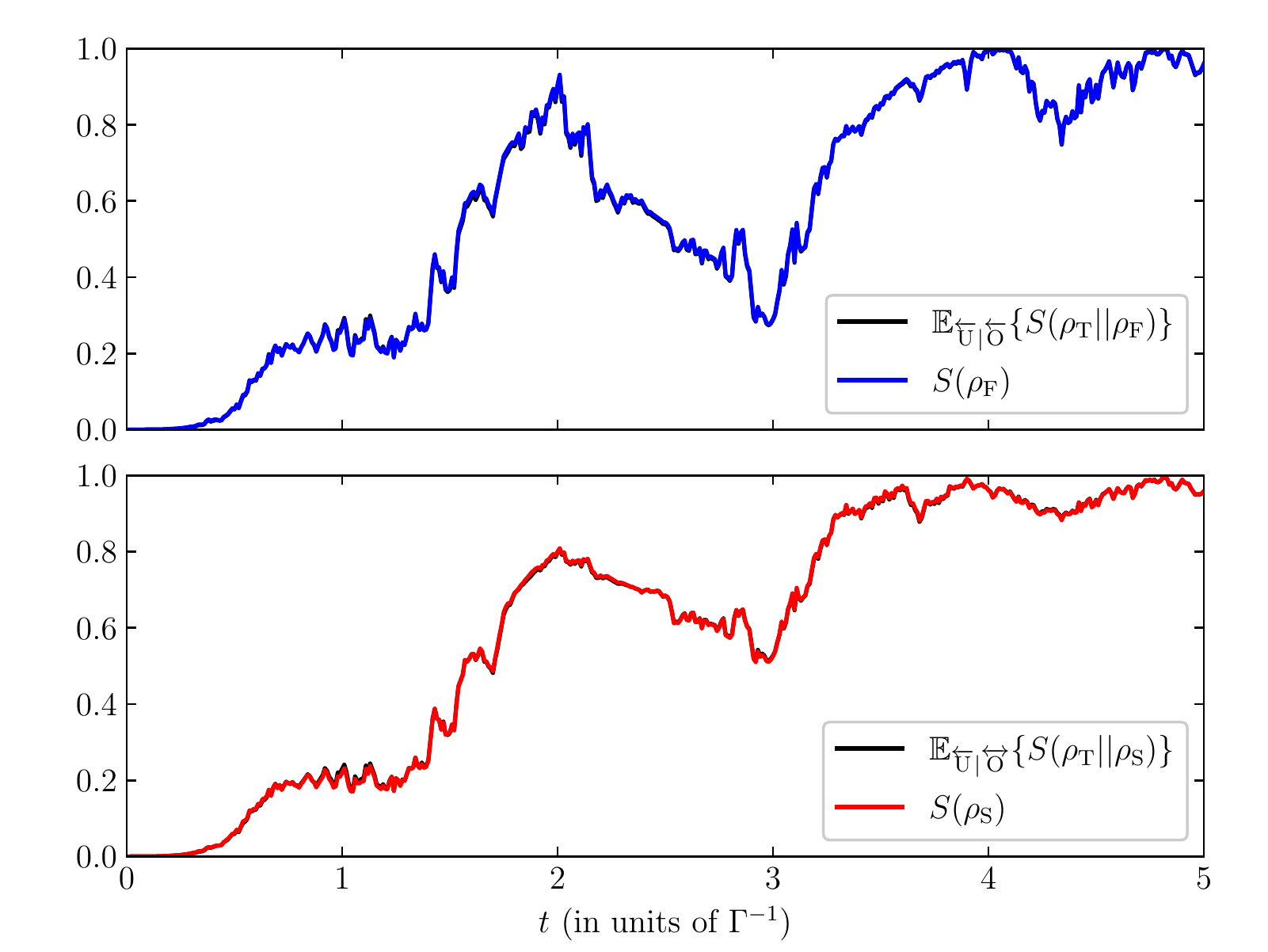}
\caption{The relative entropy risk function for the filtered state (top) and smoothed state (bottom) for the 
quantum system described in \erf{SME-God}. We have also computed the von Neumann entropy 
to illustrate the equality in \erf{RE=S} holds, using $\log_2$ in both cases. Other details, including the randomly 
generated observed record $\protect\both{\rm O}$, are as in Fig.~\ref{Fig1}~(b).}
\label{Fig2}
\end{figure}

In a similar vein to the TrSD case, when we restrict to pure true states, the expected 
risk function of the conditioned state simplifies to the von Neumann entropy of $\rho\c$, that is 
\beq
{\cal R}\c^{\rm RE}(\rho\c) = S(\rho\c)\,.
\eeq 
This follows from the fact 
that for a pure state $\rho$, $\Tr[\rho\log\rho] = 0$. {\blk Note, a similar equality was also derived in 
Ref.~\cite{GarCam19}, however they only considered the average over both the observed and unobserved records and not the conditional averages.} We verify that this equality is correct using the the previous 
physical example in \erf{SME-God}, with good agreement seen in Fig.~\ref{Fig2}. 

\section{Cost Function: Linear Infidelity}
Given the previous two cost functions, one might assume that the conditioned state $\rho\c$ would be the 
optimal estimator for any cost function that was a distance/distinguishability measure. However, this is not 
the case. The last measure we will consider as a cost function is the linear infidelity (LI) with the true state, 
i.e., ${\cal C}(\estrho,\rho\god) = 1 - L(\estrho,\rho\god)$, where the linear fidelity is defined in \erf{LF}. Note, one 
will obtain the same cost function for the Jozsa 
infidelity when assuming a pure true state, as discussed above. {\blk Once again, the task is to find the 
state that minimizes the risk function, here
\beq
{\cal R}^{\rm LI}\c(\estrho) = \mathbb{E}_{\past{\rm U}|{\rm O}_\Xi}\{1 - L(\estrho,\rho\god)\}\,.
\eeq
By the linearity of the trace, we can immediately simplify this risk function to 
${\cal R}^{\rm LI}\c(\estrho) = 1- \Tr[\estrho\rho\c]$. Furthermore, we can reframe the optimization problem in 
this case to find the estimated state $\estrho$ that maximize the linear fidelity 
$L(\estrho,\rho\c) = \Tr[\estrho\rho\c]$. 

Now, using the fact that $\rho\c$ is Hermitian and hence diagonalizable with some unitary matrix $U$, the linear 
fidelity becomes
\beq
\begin{split}
\Tr[\estrho\rho\c] =& \Tr[\estrho U \Lambda\c U\dg] = \sum_i \lambda_i (U\dg\estrho U)_{ii} \\
=& \lambda\c^{\rm max} \sum_i \frac{\lambda_i}{\lambda\c^{\rm max}}(U\dg\estrho U)_{ii}\,,
\end{split}
\eeq
where $\Lambda\c$ is a diagonal matrix whose entries are the eigenvalues $\lambda_i$ of $\rho\c$ and 
$\lambda\c^{\rm max}$ is the largest eigenvalue of $\rho\c$. Since $\rho\c$ is positive semidefinite, as it is a 
valid quantum state, $0 \leq \lambda_i/\lambda\c^{\rm max} \leq 1$. Furthermore, we know $(U\dg\estrho U)_{ii}$ is 
positive since $\estrho$ must be a valid quantum state and hence is positive semidefinite. Thus we can take the 
upper bound on $\lambda_i/\lambda\c^{\rm max}$, giving an upper bound on the linear infidelity
\beq\label{TrBound}
\Tr[\estrho\rho\c] \leq \lambda\c^{\rm max} \sum_i(U\dg \estrho U)_{ii} = \lambda\c^{\rm max}\Tr[U\dg\estrho U] = 
\lambda\c^{\rm max}\,,
\eeq
where the final equality is obtained using the cyclic property of the trace and $\Tr[\estrho] = 1$. 
We can saturate the upper bound in \erf{TrBound} by choosing the estimate state to be
\beq
\rho\c^{\rm L} = \ket{\psi\c^{\rm max}}\bra{\psi\c^{\rm max}}\,,
\eeq
$\ket{\psi\c^{\rm max}}$ is the eigenstate corresponding to the largest eigenvalue $\lambda\c^{\rm max}$ of the conditioned state. Since this state estimator is, in some sense, a purification of the conditioned state, we will call this state the {\em lustrated} conditioned state, hence the superscript ${\rm L}$.}

\label{Sec-LI}
\begin{figure}[t!]
\includegraphics[scale = 0.53]{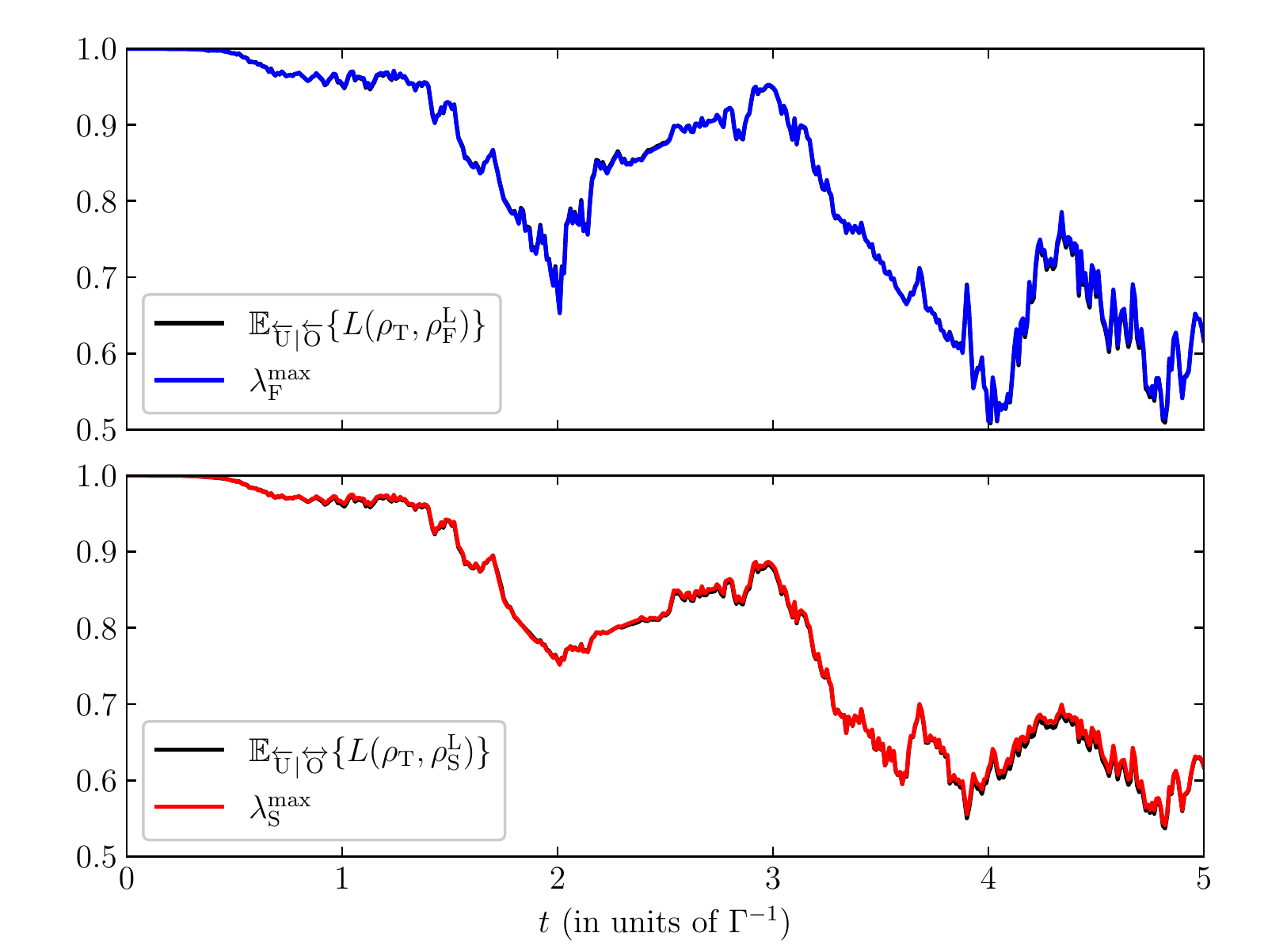}
\caption{The ensemble average of the linear fidelity between the true state and the lustrated conditioned state and 
the maximum eigenvalue of the conditioned state for the quantum system in \erf{SME-God} with Alice and Bob 
using Y-homodyne detection and photodetection, respectively. Other details, including the randomly generated 
observed record $\protect\both{\rm O}$, are as in Fig.~\ref{Fig1}~(b).}
\label{Fig3}
\end{figure}

\begin{figure*}[t!]
\includegraphics[scale = 0.545]{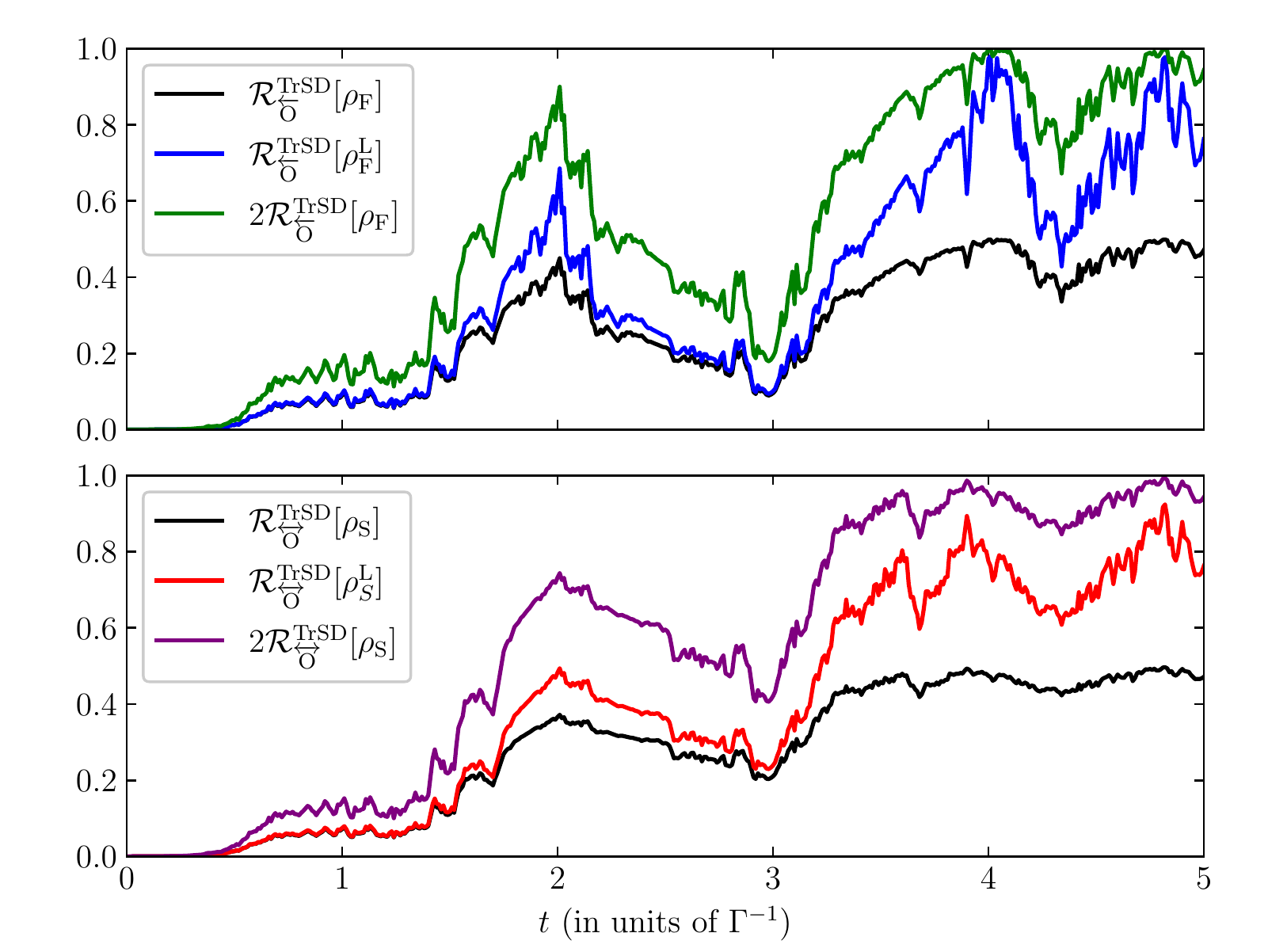}
\includegraphics[scale = 0.545]{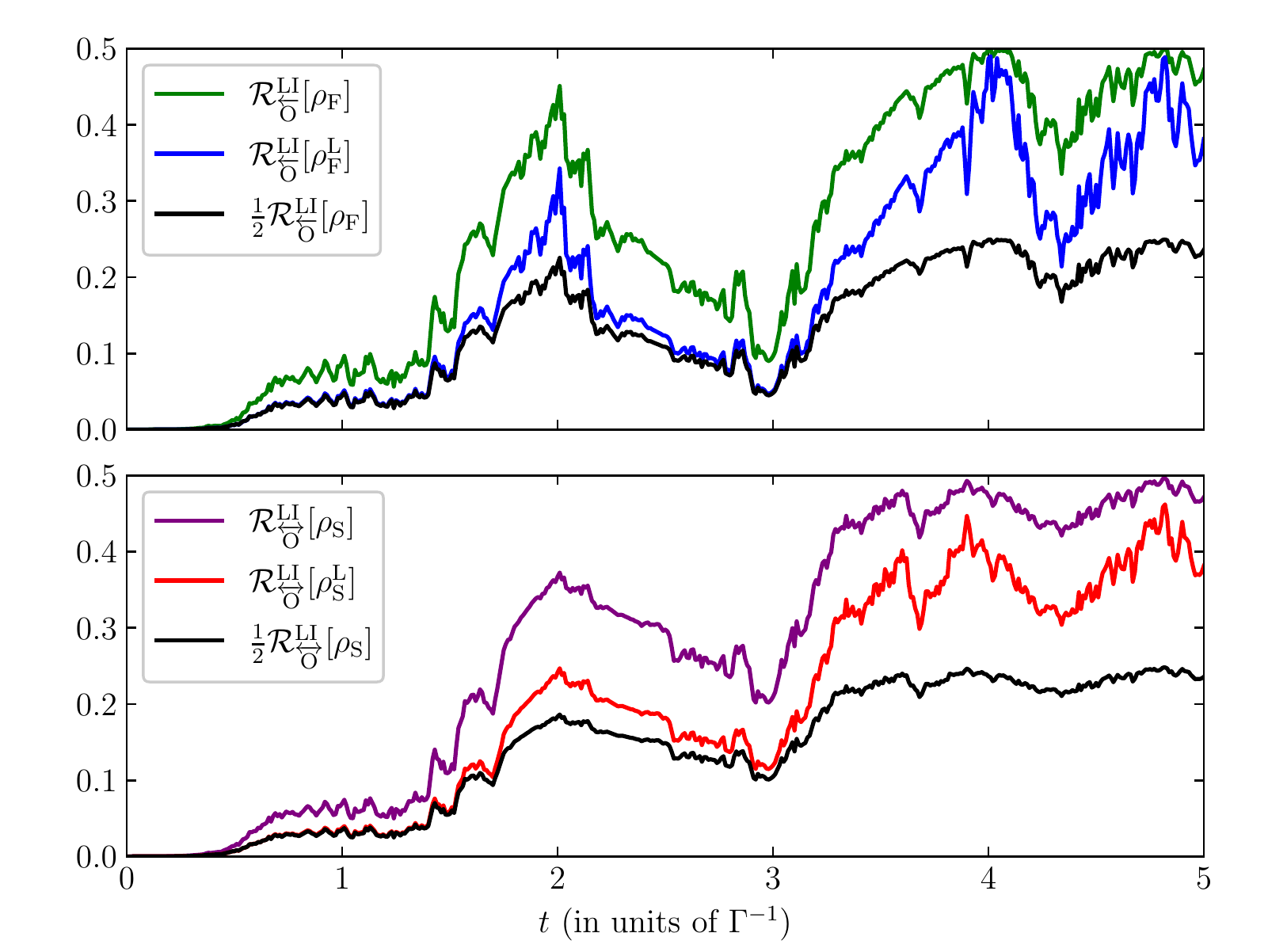}
\caption{The TrSD (left) and LI (right) risk functions for the lustrated conditioned state for the system in 
\erf{SME-God}. In both instances, we evaluate the risk function for the lustrated smoothed (top) and the lustrated 
filtered (bottom) state, where the bounds on the risk functions in terms of the relevant conditioned states are 
shown in each case, validating the bounds \erf{Bound_TrSD} (left) and \erf{Bound_LI} (right). Other details, 
including the randomly generated observed record $\protect\both{\rm O}$, are as in Fig.~\ref{Fig1}~(b).}
\label{Fig4}
\end{figure*}

As was the case for the previous two cost functions, the risk function for the optimal estimator can be simplified 
to 
\beq
{\cal R}^{\rm LI}\c[\rho\c^{\rm L}] = 1- \lambda\c^{\rm max}\,,
\eeq
which follows trivially from
\beq\label{F=eig}
{\mathbb E}_{\past{\rm U}|{\rm O}_\Xi}\{L(\rho\c^{\rm L},\rho\god)\} = \lambda\c^{\rm max}
\eeq 
In this case, contrary to the other distance measures, we 
notice that the risk function does not reduce to a simple measure of the optimal estimator. Instead the risk 
function depends on the conditioned state. To verify \erf{F=eig}, we will consider the (not-lustrated) physical system 
presented in Sec.~\ref{Sec-TrSD}. In Fig.~\ref{Fig3}, we indeed see that the average fidelity of the lustrated 
conditioned state $\rho\c\lust$ is equal to the largest eigenvalue of the conditioned state $\rho\c$.

Due to the similarity between the LI and the TrSD cost functions, it is possible to derive upper and 
lower bounds on the risk function of the lustrated state for both a LI and a TrSD cost. To begin, since the 
lustrated conditioned state is the optimal estimator for the LI risk function, we have the trivial bound 
\beq\label{triv-LI}
{\cal R}^{\rm LI}\c[\rho\c\lust]\leq {\cal R}^{\rm LI}\c[\rho\c]\,,
\eeq 
where an equality occurs when the conditioned state is pure. 
Whilst this is trivial for a LI cost, this relationship places a non-trivial upper bound on the TrSD risk function 
for the lustrated conditioned state. Specifically, it is easy to show, using 
\erf{F=P} and assuming a pure true state, that \erf{triv-LI} implies that
${\cal R}^{\rm TrSD}\c[\rho\c\lust] \leq 2{\cal R}^{\rm TrSD}\c[\rho\c]$.
Similarly, we can derive a lower bound on the LI risk function for the lustrated conditioned state by 
considering ${\cal R}^{\rm TrSD}\c[\rho\c] \leq {\cal R}^{\rm TrSD}\c[\rho\c\lust]$. In this case, we obtain the 
lower bound $\half{\cal R}^{\rm LI}\c[\rho\c]\leq {\cal R}^{\rm LI}\c[\rho\c\lust]$. As a result, we obtain the 
following bounds for the on the risk functions of the lustrated conditioned state
\begin{align}
&{\cal R}^{\rm TrSD}\c[\rho\c] \leq {\cal R}^{\rm TrSD}\c[\rho\c\lust] \leq 2{\cal R}^{\rm TrSD}\c[\rho\c]\,.
\label{Bound_TrSD}\\
&\half{\cal R}^{\rm LI}\c[\rho\c] \leq {\cal R}^{\rm LI}\c[\rho\c\lust] \leq {\cal R}^{\rm LI}\c[\rho\c]\,,
\label{Bound_LI}
\end{align}
To verify these bounds on the risk functions, we will once again consider the physical example presented 
in Sec.~\ref{Sec-TrSD}. In Fig.~\ref{Fig4}, we consider the both TrSD (left) and LI (right) risk functions for 
the smoothed (top) and filtered (bottom) state, observing the bounds in \erfs{Bound_LI}{Bound_TrSD}. {\blk Note, 
one might think that the upper bound in \erf{Bound_TrSD} could become a trivial bound when 
${\cal R}^{\rm TrSD}\c[\rho\c] > 1$ as the maximum value of the TrSD risk function is $2$. In fact, this is never trivial 
as, from \erf{simp_risk}, we can see that ${\cal R}^{\rm TrSD}\c[\rho\c] \leq 1$. }

\section{Conclusion}
In this paper, we have shown that the smoothed quantum state is an 
optimal state estimator. Specifically, the smoothed quantum state simultaneously minimizes the risk function for 
a trace-square deviation and relative entropy cost function. Furthermore, we showed that, in both 
cases, the risk function of the smoothed state reduced to simple measures acting the solely on the 
smoothed state, specifically, the impurity and von Neumann entropy of the smoothed state, respectively. 
However, when we considered the linear infidelity as a cost function we found, somewhat counter-intuitively, that 
the smoothed quantum state was not optimal. Instead, the the lustrated smoothed state, defined as the eigenstate 
corresponding to the maximum eigenvalue of the smoothed quantum state, is the optimal estimator for such a cost 
function. 

As was the case for the other cost functions, we showed that the linear infidelity risk function of the lustrated 
smoothed state reduces to a simple measure. However, in this case the measure does not depend on the 
lustrated smoothed state itself, rather it depends on the maximum eigenvalue of the smoothed quantum state. 
Finally, we calculated some upper and lower bounds on the the risk function of the lustrated smoothed 
state, for both the trace-square deviation and the linear infidelity. 

Since the lustrated state is pure, an obvious question is whether it is related to the pure states in the most likely 
paths approach of Refs.~\cite{Cha13, Web14}. This, and many other related questions, are answered by the 
general cost function approach to quantum state estimation using past and future measurement records introduced 
in Ref.~\cite{CGLW20}. However, the most-likely path \cite{Cha13, Web14, CGLW20} is restricted to homodyne-like 
unknown records, whereas in this paper we have used an example here the unknown record is comprised of 
discrete photon counts. It remains an open question as to whether the most-likely path cost functions of 
Ref.~\cite{CGLW20} can be generalized to such cases.

\begin{acknowledgements}
We would like to thank Areeya Chantasri for many useful discussions regarding this work.
We acknowledge the traditional owners of the land on which this work was undertaken 
at Griffith University, the Yuggera people. This research is funded by the Australian Research Council 
Centre of Excellence Program CE170100012. K.T.L. is supported by an Australian Government 
Research Training Program (RTP) Scholarship.
\end{acknowledgements}

\appendix
\section{Numerics}\label{App-code}
{\blk In this appendix, we present the methods used to compute the filtered, true and smoothed quantum 
states and all the measures in this paper. To begin, a typical homodyne measurement current, which will remain fixed for all calculations, was generated in 
with parallel the associated filtered state and calculating the measurement current via 
\beq
\dd J_\phi(t) = \Tr[\hat{c}_\phi\rho\fil(t) + \rho\fil(t)\hat{c}_\phi\dg] + \dd W\fil(t)\,,
\eeq
where $\dd W\fil$ is the filtered 
innovation generated from a Gaussian distribution with the moments in \erf{innovation}. The filtered state was 
computed using quantum maps \cite{WisMil10}, where the evolution of the filtered state in a finite time step 
$\delta t$ is given by
\beq\label{norm-filt}
\rho\fil(t + \delta t) = \frac{{\cal M}_H{\cal M}_{\dd J_\phi(t)}{\cal M}_{u} \rho\fil(t)}{\Tr[{\cal M}_H
{\cal M}_{\dd J_\phi(t)}
{\cal M}_u \rho\fil(t)]}\,,
\eeq
where the completely positive map ${\cal M}_A$ subscripts denote the particular type of 
evolution the system is undergoing: $\hat{H}$ denotes the Hamiltonian part, $\dd J_\phi(t)$ denotes the homodyne part 
and $u$ denotes the remaining unconditioned dynamics. The Hamiltonian part is 
${\cal M}_H \bullet= {\rm exp}(-iH\delta t) \bullet {\rm exp}(iH\delta t)$. The unconditioned map can be described  
by averaging over Bob's jump process to make a trace-preserving map, 
\beq
{\cal M}_u \bullet = \sum_{\dd N(t) = 0}^{1}\hat{M}_{\dd N(t)} \bullet \hat{M}_{\dd N(t)}\dg\,.
\eeq
The homodyne (conditioned) map is described by a single measurement operator 
${\cal M}_{\dd J_\phi(t)} \bullet = \hat{M}_{\dd J_\phi(t)}\bullet\hat{M}_{\dd J_\phi(t)}\dg$.
In particular, we used completely positive quantum maps 
\cite{GueWis20}, where the $\hat{M}$ operators have been taken to a second order in $\delta t$ to ensure the 
positivity of the quantum state to high accuracy. For details on the particular operators used for the homodyne 
measurement and the jump measurement see Ref.~\cite{GueWis20}. 

With this typical record, both the unnormalized filtered state $\tilde{\rho}\fil$ and the unnormalized true state 
$\tilde{\rho}\god$  can be computed. The unnormalized filtered state is computed using 
\erf{norm-filt} without the trace term on the denominator, and the unnormalized true state evolves as
\beq
\tilde\rho\god(t + \delta t) = {\cal M}_H {\cal M}_{\dd J_\phi(t)} {\cal M}_{\dd N(t)} \rho\god(t)\,.
\eeq
The reason why the 
unnormalized versions of these states are computed, as opposed to the normalized version, is is because their 
traces correspond to the ostensible probability distributions $\Tr[\tilde{\rho}\fil] = \wp_{\rm ost}(\past{\rm O})$ and 
$\Tr[\tilde{\rho}\god] = \wp_{\rm ost}(\past{\rm O},\past{\rm U})$. These distributions are needed for computing the 
ensemble average over $\past{\rm U}$ given $\past{\rm O}$ via $\wp_{\rm ost}(\past{\rm U}|\past{\rm O}) = 
\wp_{\rm ost}(\past{\rm O},\past{\rm U})/\wp_{\rm ost}(\past{\rm O})$. Specifically, the ensemble average using this 
conditional probability is
\beq
\mathbb{E}_{\past{\rm U}|\past{\rm O}}\{\bullet\} = \frac{1}{N_{\rm tot}}\sum_{n = 1}^{N_{\rm tot}}\frac{\Tr[\tilde{\rho}\god^{(n)}]}
{\Tr[\tilde{\rho}\fil]} \bullet\,,
\eeq
where the superscript $(n)$ labels the $n$th realization of the true state with $N_{\rm tot}$ being the total number 
of realizations computed.

For the smoothed quantum state, it is necessary to perform the ensemble average 
$\mathbb{E}_{\past{\rm U}|\both{\rm O}}$. Thus we require the ostensible distribution $\wp_{\rm ost}(\past{\rm U}|
\both{\rm O}) = \wp_{\rm ost}(\both{\rm O},\past{\rm U})/\wp_{\rm ost}(\both{\rm O})$. Both the numerator and 
denominator are obtained by introducing the retrofiltered effect $\hat{E}\rfil$, a POVM element that evolve 
backwards-in-time from a final uninformative state $\hat{E}\rfil(T) = I$ conditioning on the measurement result back 
to the time $\tau$. The retrofiltered effect is computed as the adjoint of the unnormalized filtered state, that is,
\beq
\hat{E}\rfil(t - \delta t) = {\cal M}_u\dg {\cal M}_{\dd J_\phi(t)}\dg {\cal M}_{H}\dg \hat{E}\rfil(t)\,,
\eeq
where $\Tr[\hat{E}\rfil\rho] = \wp(\fut{\rm O}|\rho)$. The ostensible distributions are then obtain by $\Tr[\tilde{\rho}
\god\hat{E}\rfil] = \wp_{\rm ost}(\both{\rm O},\past{\rm U})$ and 
$\Tr[\tilde{\rho}\fil\hat{E}\rfil] = \wp_{\rm ost}(\both{\rm O})$. Thus the ensemble average conditioning on 
$\both{\rm O}$ is computed as
\beq
\mathbb{E}_{\past{\rm U}|\both{\rm O}}\{\bullet\} = \frac{1}{N_{\rm tot}}\sum_{n = 1}^{N_{\rm tot}}\frac{\Tr[\tilde{\rho}
\god^{(n)} \hat{E}\rfil]}{\Tr[\tilde{\rho}\fil\hat{E}\rfil]} \bullet\,.
\eeq
{\blk Note, for a fair comparison of the various equalities presented in this paper, the ensemble of true states (of 
size $M$) used to compute the ensemble average for the smoothed state were generated independently of the 
ensemble of true states (of size $N$) used to compute the ensemble averages of other quantities, like the linear 
fidelities and relative entropies.}

\end{document}